\documentclass[twocolumn,english,prb,superscriptaddress,showpacs,amsmath,amssymb]{revtex4}
\usepackage{graphicx,esint,bm,float,lipsum,babel}
\makeatletter
\@ifundefined{textcolor}{}
{%
 \definecolor{BLACK}{gray}{0}
 \definecolor{WHITE}{gray}{1}
 \definecolor{RED}{rgb}{1,0,0}
 \definecolor{GREEN}{rgb}{0,1,0}
 \definecolor{BLUE}{rgb}{0,0,1}
 \definecolor{CYAN}{cmyk}{1,0,0,0}
 \definecolor{MAGENTA}{cmyk}{0,1,0,0}
 \definecolor{YELLOW}{cmyk}{0,0,1,0}
 }
\makeatother

\begin{document}

\title{Double resonance Raman modes in mono- and few-layer MoTe$_2$}

\author{Huaihong Guo}
\email{hhguo@alum.imr.ac.cn}
\affiliation{Department of Physics, Tohoku University, Sendai, Miyagi 980-8578, Japan}
\author{Teng Yang}
\email{yangteng@imr.ac.cn}
\affiliation{Shenyang National Laboratory for Materials Science, Institute of Metal Research,
Chinese Academy of Sciences, 72 Wenhua Road, Shenyang 110016, China}
\affiliation{Department of Physics, Tohoku University, Sendai, Miyagi 980-8578, Japan}
\author{Mahito Yamamoto}
\affiliation{International Center for Materials Nanoarchitectonics (WPI-MANA), National
Institute for Materials Science (NIMS), Tsukuba, Ibaraki 305-0044, Japan}
\author{Lin Zhou}
\affiliation{Department of Electrical Engineering and Computer Science, Massachusetts Institute
of Technology,Cambridge, Massachusetts 02139-4037, USA}
\affiliation{Department of Physics, Massachusetts Institute of Technology, Cambridge,
Massachusetts 02139-4307, USA}
\author{Ryo Ishikawa}
\affiliation{Department of Functional Materials Science, Graduate School of Science and
Engineering, Saitama University, Saitama 338-8570, Japan}
\author{Keiji Ueno}
\affiliation{Department of Chemistry, Graduate School of Science and Engineering, Saitama
University, Saitama 338-8570, Japan}
\author{Kazuhito Tsukagoshi}
\affiliation{International Center for Materials Nanoarchitectonics (WPI-MANA), National
Institute for Materials Science (NIMS), Tsukuba, Ibaraki 305-0044, Japan}
\author{Zhidong Zhang}
\affiliation{Shenyang National Laboratory for Materials Science, Institute of Metal Research,
Chinese Academy of Sciences, 72 Wenhua Road, Shenyang 110016, China}
\author{Mildred~S.~Dresselhaus}
\affiliation{Department of Electrical Engineering and Computer Science, Massachusetts
Institute of Technology,Cambridge, Massachusetts 02139-4037, USA}
\affiliation{Department of Physics, Massachusetts Institute of Technology, Cambridge,
Massachusetts 02139-4307, USA}
\author{Riichiro~Saito}
\affiliation{Department of Physics, Tohoku University, Sendai, Miyagi 980-8578, Japan}

\date{\today}

\begin{abstract}
We study the second-order Raman process of mono- and few-layer MoTe$_2$, by combining
{\em ab initio} density functional perturbation calculations with experimental Raman
spectroscopy using 532, 633 and 785 nm excitation lasers. The calculated electronic band
structure and the density of states show that the resonance Raman process occurs at the M
point in the Brillouin zone, where a strong optical absorption occurs due to a logarithmic
Van-Hove singularity of electronic density of states. Double resonance Raman process with
inter-valley electron-phonon coupling connects two of the three inequivalent M points in
the Brillouin zone, giving rise to second-order Raman peaks due to the M point phonons.
The calculated vibrational frequencies of the second-order Raman spectra agree with the
observed laser-energy dependent Raman shifts in the experiment.
\end{abstract}

\maketitle

\section{Introduction}

Transition metal dichalcogenides (TMDs) have been intensively studied over past decades, and
recently attracted tremendous attention for a number of applications\cite{Radisavljevic11,Wang12,
Lopez-Sanchez13,Lin14,Pradhan14}, since TMDs have a finite energy band gap, which is more
advantageous than graphene. Among the TMDs, MoTe$_2$ has a stronger spin-orbit coupling than other
Mo related TMD materials (MoX$_2$, X = S, Se), which leads to a longer decoherence time both for
valley and spin freedom \cite{Pradhan14} and thus MoTe$_2$ is considered to be used for ideal
valleytronic devices\cite{Xiao12,Cao12}. Unlike other TMDs, MoTe$_2$ has unique structural properties,
such as a thermally-induced structural phase transition at around 900$^\circ$C from the $\alpha$-phase
diamagnetic semiconductor to the $\beta$-phase paramagnetic metal\cite{Vellinga70, Duerloo14}.
MoTe$_2$ has the smallest direct energy band gap ($\sim$ 1.1 eV) at the K point (hexagonal
corner) in the Brillouin zone (BZ) among all semiconducting TMDs, the band gap is similar to the
indirect gap ($\sim$ 1.1 eV) of silicon\cite{Streetman00,Ruppert14}. MoTe$_2$ is thus suitable for
nanoelectronics devices\cite{Lin14,Pradhan14} and for studying exciton effects at the K points\cite{Wilson69}.
Raman spectroscopy is a useful tool for characterizing TMDs since we get many information
\cite{Lee10,Sahin13,Li12} such as number of layers, electronic and phonon properties of TMDs.
Recent Raman spectroscopy studies of monolayer and few-layer MoTe$_2$ using visible light (1.9
$\sim$ 2.5 eV) show that there exists some unassigned Raman spectra which can not be assigned
to a first-order Raman process\cite{Pradhan14,yamamoto14}, and the assignment of these Raman
spectra is the subject of the present paper.

Raman spectra of TMDs have been reported for several decades\cite{Wieting71,Chen74,Wakabayashi75,
Mead77,Beal79,Sekine80a,Sekine80b,Sugai81,Sekine84,Stacy85,Sourisseau89,Sourisseau1991,Frey99}.
The first-order Raman process in TMDs has been widely investigated and is now well understood
\cite{Sugai81,Sugai82,Sekine80a,Beal79,Sanchez11,Ding11,Luo13a,Luo13b,yamamoto14}. However, the
study of second-order Raman spectra did not give consistent assignments for different TMDs\cite{Stacy85,
Chen74,Sekine84,Chakraborty12,Zhao13,Terrones14}. For example, Stacy \textit{et al.}\cite{Stacy85}
and many other groups\cite{Chen74,Feldman96,Frey99,Windom11,Li12,Chakraborty13} assigned some of
the Raman peaks of 2H-MoS$_2$ to either combination or overtone second-order phonon modes occurring
at the M point (center of hexagonal edge) in the Brillouin zone. Terrones \textit{et al.} \cite{Terrones14}
also assigned phonon modes in few-layer WSe$_2$ according to the M point phonon. On the other hand,
Sourisseau \textit{et al.} \cite{Sourisseau89,Sourisseau1991} ascribed the second-order Raman
process of 2H-WS$_2$ to the K point phonons. Berkdemir \textit{et al.}\cite{Berkdemir13} considered
that electrons are scattered by the M point phonons between the K point and I point (one point along
$\Gamma$K direction), and that this process is responsible for the double resonant Raman process in
few-layer WS$_2$. Furthermore, Sekine \textit{et al.}\cite{Sekine84} found a dispersive Raman peak
in 2H-MoS$_2$ and interpreted this peak in terms of a two-phonon Raman process at the K point. The
different assignments for both few-layer or bulk MoS$_2$ and WS$_2$, either at the M or K points, have
not been well investigated, since most of these assignments to second-order Raman peaks have been made
merely by comparing phonon frequencies with reference to some inelastic neutron scattering results\cite{Wakabayashi75,Sourisseau1991}.
Thus a detailed analysis with use of double resonance Raman theory is needed, especially for understanding
both mono- and few-layer MoTe$_2$ that are discussed in this paper.

In the present paper, we mainly study the second-order Raman spectra of monolayer MoTe$_2$,
based on {\em ab initio} electronic band calculations, density functional perturbation theory and
associated experimental Raman spectroscopy. From our electronic band structure calculation, we have
found a saddle point of the energy dispersion at the M points in the BZ, which gives rise
to two-dimensional Van Hove singularity (VHS) in the electronic density of states. The energy separation
between two VHSs (2.30 and 2.07 eV) are matched to the 2.33 eV (532 nm) and 1.96 eV (633 nm) laser
excitations in our experiment in which resonant optical absorption occurs at the M points. Consequently, inter-valley
electron-phonon scattering process occurs between two of the three inequivalent M (M, M$'$, M$''$) points,
giving rise to a double resonance Raman process. We propose that this double resonance process occurring at the
M points is essential for the interpretation of the observed second-order Raman peaks of MoTe$_2$. The
experimental second-order Raman frequencies as a function of laser excitation energy confirm our
interpretation of the double resonance Raman spectra.

Organization of the paper is as follows. In II and III, we briefly show computational details and
experimental methods, respectively. In IV, both experimental and theoretical results for mono- and few-layer
MoTe$_2$ Raman spectra are given. We assign second-order Raman peaks by comparing the calculated results with the
observed Raman spectra. In V summary of the paper is given. Further information such as layer-number dependence
is discussed in Supplemental Material.

\section{Computational details}
We performed the electronic and phonon energy dispersion calculations on monolayer and few-layer
MoTe$_2$ by using first-principles density functional theory within the local density approximation
(LDA) as implemented in the Quantum-Espresso code \cite{Giannozzi09}. The structures of monolayer
and few-layer MoTe$_2$ that are used in the present experiment are $\alpha$-MoTe$_2$\cite{yamamoto14}.
The AA$'$ and AA$'$A stacking geometries are used in the computational study for bi-layer and tri-layer
MoTe$_2$, respectively, since these stacking geometries are more stable than the other possibilities\cite{Tao14}. AA$'$
stacking refers to the geometry which has a Mo (Te) atom of one layer on top of a Te (Mo) atom of
the other layer. The few-layer MoTe$_2$ are separated by 25 \AA\ from one another in a unit cell of
the calculation to eliminate the inter-few-layer interaction. We used projector augmented-wave (PAW)
pseudopotentials\cite{blochl94} with a plane-wave cutoff energy of 65 Ry to describe the interaction
between electrons and ions. The spin-orbit split electronic band structures were calculated by
relativistic pseudopotentials derived from an atomic Dirac-like equation. The atomic coordinates are
relaxed until the atomic force is less than 10$^{-5}$ Ry/Bohr. The lattice parameters used in this paper
are listed in Table S1 and agree well with the experiment and other calculations \cite{Ding11,Puotinen61,
Kang13}. The Monkhorst-Pack scheme \cite{Monkhorst76} is used to sample the Brillouin zone (BZ) over a
17$\times$17$\times$1 $\textit{k}$-mesh for MoTe$_2$. The phonon energy dispersion relations of MoTe$_2$
are calculated based on density functional perturbation theory \cite{Baroni01}. In order to visualize
the vibrational modes at the M point, we use a 2$\times$2$\times$1 supercell in which the M point is
zone-folded to the $\Gamma$ point in the folded BZ.

The optical absorption spectrum is calculated by the real ($\epsilon$$'$) and imaginary ($\epsilon$$''$)
parts of the dielectric function as a function of photon energy, respectively, based on the PAW
methodology\cite{Gajdos06} and the conventional Kramers-Kronig transformation. The absorption coefficient
$\alpha$ is described by $\alpha$ = 4$\pi\kappa$E$_\gamma$/($\textit{hc}$), where E$_\gamma$ is the
incident laser excitation energy, $\textit{h}$ the Planck constant, $\textit{c}$ the speed of light, and $\kappa$ is
the extinction coefficient\cite{Zimmermann00}, that is, $\kappa$ = $\sqrt{(\sqrt{\epsilon'^2 +
\epsilon''^2} - \epsilon')/2}$.

\begin{figure*}[t]
\includegraphics[width=1.60\columnwidth]{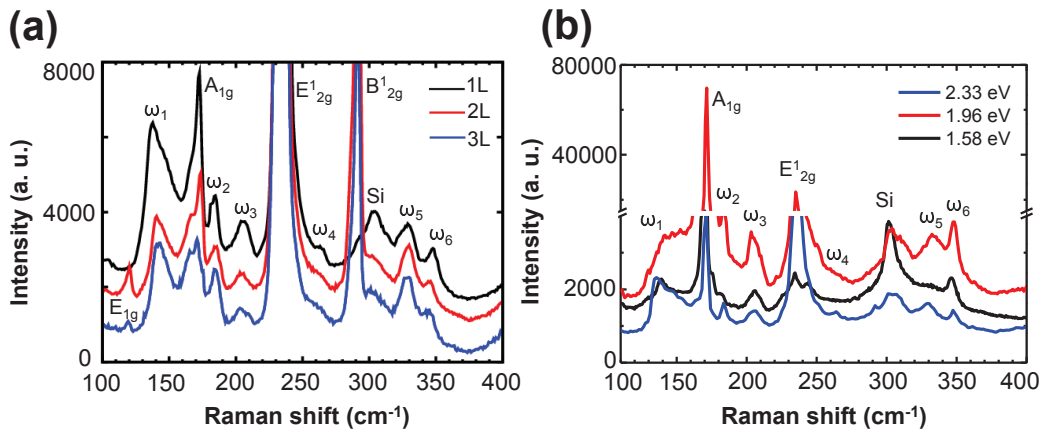}
\caption{(Color online) (a) Raman spectra of mono-, bi-, and tri-layer MoTe$_2$ under 2.33 eV
(532 nm) laser excitation energy, showing the small intensity peaks $\omega_i$ (i = 1, 2, ..., 6)
and an additional calibration peak from Si besides the typical Raman peaks of A$_{1g}$, E$_{2g}^1$
and B$_{2g}^1$ that are frequently observed in MoTe$_2$\cite{Pradhan14,yamamoto14}. (b) Raman spectra of monolayer MoTe$_2$
under 2.33 eV (532 nm), 1.96 eV (633 nm), and 1.58 eV (785 nm) laser excitation energies.}
\label{Fig1}
\end{figure*}

\section{Experimental method}
Bulk crystals of $\alpha$-MoTe$_2$ were prepared through a chemical vapor transport method\cite{Lieth77}.
Atomically thin crystals of MoTe$_2$ were mechanically exfoliated from the bulk crystals onto silicon
substrates with a 90 nm-thick oxide layer on top. The thicknesses of the atomic crystals were identified using
optical microscopy and first-order Raman spectroscopy \cite{yamamoto14}. Raman spectroscopy measurements
for monolayer MoTe$_2$ were performed using 532, 633 and 785 nm excitation lasers for discussing the observed phonon
dispersion due to double resonance Raman theory\cite{Saito01}. Raman spectroscopy studies of few-layer MoTe$_2$
under 532 nm laser excitation were also performed and the results are discussed in the supporting materials.
The grating sizes were 1200, 1800 and 2400 lines/mm for the 785, 633 and 532 nm laser excitation measurements,
respectively. The magnification of the objective lens was 100x. The accumulation times were 150-200 seconds.
The laser power was kept below 0.1 mW, 0.2 mW and 3.0 mW for the 532, 633 and 785 nm excitation lasers,
respectively, in order to avoid damage to the samples. All measurements were performed at room temperature in
the backscattering configuration. Since this paper gives greater emphasis to a theoretical understanding
of the 2nd-order Raman scattering processes, we do not present many experimental details. However, more
details about the characterization of the number of layers in our MoTe$_2$ samples can be obtained in one of
the previous works on MoTe$_2$ synthesis and characterization, based on the first-order Raman spectra\cite{yamamoto14}.

\section{Results and discussion}
Figure~\ref{Fig1}(a) shows Raman spectra of mono- to tri-layer MoTe$_2$ under the 532 nm (2.33 eV) laser excitation.
The Raman spectra show three strong peaks, including the in-plane E$^1_{2g}$ mode and the out-of-plane A$_{1g}$ mode
for monolayer and few-layer MoTe$_2$, as well as the E$_{1g}$ mode and the B$^1_{2g}$ mode for bi- and tri-layer MoTe$_2$,
that were observed previously\cite{Pradhan14,yamamoto14,note1}. These spectra are assigned to the first-order Raman spectra
of the $\Gamma$ point phonons\cite{yamamoto14}. However, we observe additional peaks with relatively small
intensities, which can not be assigned as first-order Raman spectra\cite{yamamoto14}. In Fig.~\ref{Fig1}(a),
we denote the weak Raman spectra as $\omega_i$ (i = 1, 2, ..., 6).

Since the second-order Raman spectra are dispersive as a function of excitation energy\cite{Saito01}, we performed
Raman spectroscopy measurements using different excitation laser energies. In Fig.~\ref{Fig1}(b) we show Raman spectra
of monolayer MoTe$_2$ under 2.33 eV (532 nm), 1.96 eV (633 nm) and 1.58 eV (785 nm) laser excitation energies. Here
the Raman intensities are normalized using the non-resonant Si peak intensity at 520 cm$^{-1}$ (not shown in Fig.~\ref{Fig1}(b)).
We find that the peak positions of each $\omega_i$ either upshift or downshift by up to several cm$^{-1}$ for different
laser excitation energies, suggesting that these peaks are associated with a second-order Raman process\cite{Saito01}.
\begin{figure*}[t]
\includegraphics[width=1.9\columnwidth]{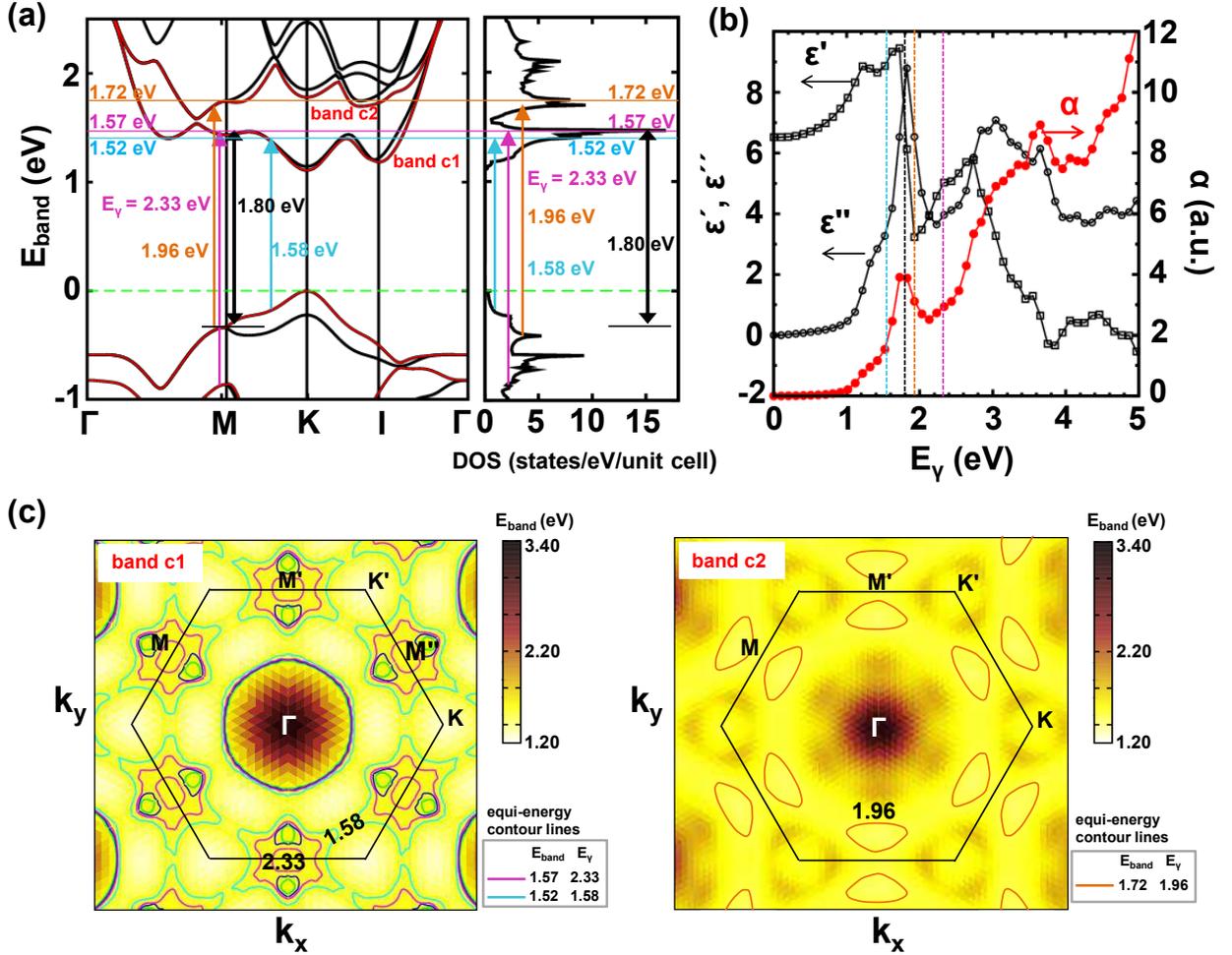}
\caption{(Color online) (a) Electronic energy band structure E(k) and the density of states (DOS), (b) the real ($\epsilon$$'$),
and imaginary ($\epsilon$$''$) parts of the dielectric function and the optical absorption coefficient $\alpha$ as a function
of photon energy E$_\gamma$, and (c) an equi-energy contour plot of the two conduction bands (c1 and c2) with the
same spin for monolayer MoTe$_2$. Energy bands are split by spin-orbit interaction, the conduction bands c1, c2 and the valence
bands which have the same spin\cite{Xiao12,Wang14} are highlighted in red solid lines in (a). When we consider energy separation
between two Van-Hove singularities (VHSs) of the DOS, the peak at 1.80 eV in the absorption spectrum corresponds to optical
transition at the M point (black double headed arrow in (a)). (c) The equi-energy contour plot for VHS E$_{band}^{c1}$ = 1.52 and
1.57 eV (left) and E$_{band}^{c2}$ = 1.72 eV (right). All equi-energy contours appear around the M point. The E$_{band}$ energies
of 1.52, 1.57 and 1.72 eV (horizontal lines in (a)) correspond to the experimental laser energies (1.58, 2.33 and 1.96 eV) in blue,
pink and orange colors.}
\label{Fig2}
\end{figure*}

Considering that the second-order process is usually connected to the electronic properties by double resonance
Raman theory\cite{Saito01}, we calculate the electronic band structure, the density of states (DOS) and the
optical absorption spectra of monolayer MoTe$_2$ as shown in Fig.~\ref{Fig2}. Our calculated energy dispersion
reproduces the main observed features\cite{Kumar12,Ding11,Ruppert14} such as the direct electronic band gap
(1.00 eV) and the spin-orbit splitting of the valence band ($\sim$230 meV) at the K point which are close to
1.10 eV and 250 meV, respectively, obtained from the optical absorption measurements\cite{Ruppert14}. It is noted that
the energy band gap is underestimated by the local density functional method. In order to fit the experimental
data, the conduction bands are upshifted by 0.10 eV, so are the DOS and optical absorption spectrum. In
Fig.~\ref{Fig2}(a), we show the calculated electronic energy band and DOS. From the parabolic bands around the
K point, two-dimensional (2D) DOS gives constant values, while at the M point we have found saddle points
of the energy dispersion in the electronic band, the corresponding DOS gives logarithmic $\log$$|${$E$ - $E_0$}$|$
divergence as is known as Van Hove Singularity (VHS) in 2D systems.

In Fig.~\ref{Fig2}(b), we show the calculated real ($\epsilon'$) and imaginary ($\epsilon''$) parts of the
dielectric constant and the optical absorption coefficient $\alpha$ as a function of photon energy. $\epsilon''$
has a nonzero value from the photon energy E$_\gamma$ = 1.10 eV which corresponds to the energy gap, and
a peak at E$_\gamma$ $\sim$ 1.80 eV. A strong optical absorption at 1.80 eV from our calculation agrees well
with the large peak at around 1.83 eV from the optical absorption measurement by Ruppert et al.\cite{Ruppert14}
on monolayer MoTe$_2$. Moreover, the electronic excitation energy of 2.30 eV (2.07 eV) from the third (top)
valence band to the lowest (third lowest) conduction band at the M point matches very well with the laser energy
2.33 eV (1.96 eV) used in our experiment, as indicated by pink (orange) arrows in the band structure of Fig.~\ref{Fig2}(a).
Thus resonant optical absorption at these laser lines is expected.



In Fig.~\ref{Fig2}(c), we plot equi-energy contour for the conduction band energies E$_{band}$ which correspond to the
laser excitation energies E$_{\gamma}$. E$_{band}$ = 1.52 and 1.57 eV for c1 and 1.72 eV for c2 correspond to the E$_{\gamma}$ =
1.58, 2.33 and 1.96 eV, respectively, as indicated by blue, pink, and orange arrows in Fig.~\ref{Fig2}(a). The conduction
bands c1, c2 and the valence bands which have the same spin\cite{Xiao12,Wang14} are highlighted in red solid lines in the
band structure. The equi-energy contour lines E$^{c1}_{band}$ and E$^{c2}_{band}$ in Fig.~\ref{Fig2}(c) both show the saddle
point shape of the energy dispersion in the vicinity of the M point.
That is, the energy band E$^{c1}_{band}$ (E$^{c2}_{band}$) starting from the M point monotonically decreases (increases) along
the M-K direction, whereas this band energy increases (decreases) along the M-$\Gamma$ direction. When we decrease laser excitation
energies from 2.33 to 1.58 eV, the contour lines in Fig.~\ref{Fig2}(c) left change the shapes around the M point, suggesting
that the area inside of the contour line for each M point decreases with increasing laser excitation energy. This situation is opposite
to the case of graphene in which the equi-energy contour is almost circular and the area increases with increasing energy.
\begin{figure}[t]
\includegraphics[width=0.80\columnwidth]{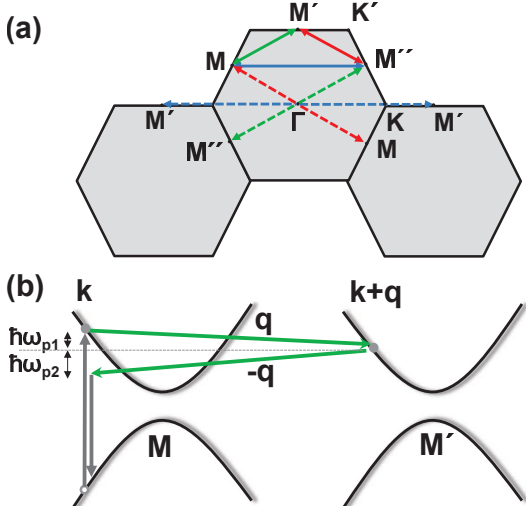}
\caption{(Color online) (a) In the double-resonance process, photon-excited electrons are scattered inelastically by emitting phonons and make inter-valley (i.e., M$\longleftrightarrow$M$'$, M$'$$\longleftrightarrow$M$''$, and M$\longleftrightarrow$M$''$) transitions, as shown by the solid double-end arrows in the first BZ. The related phonon vectors $\vec q$ measured from the $\Gamma$ point are also given by the six dashed arrows. (b) The schematics of the double resonance process (electron-photon and electron-phonon processes). This process generates two phonons ($\hbar \omega_{p1}$, -$\vec q$) and ($\hbar \omega_{p2}$, $\vec q$).}
\label{Fig3}
\end{figure}

\begin{figure*}[t]
\includegraphics[width=1.6\columnwidth]{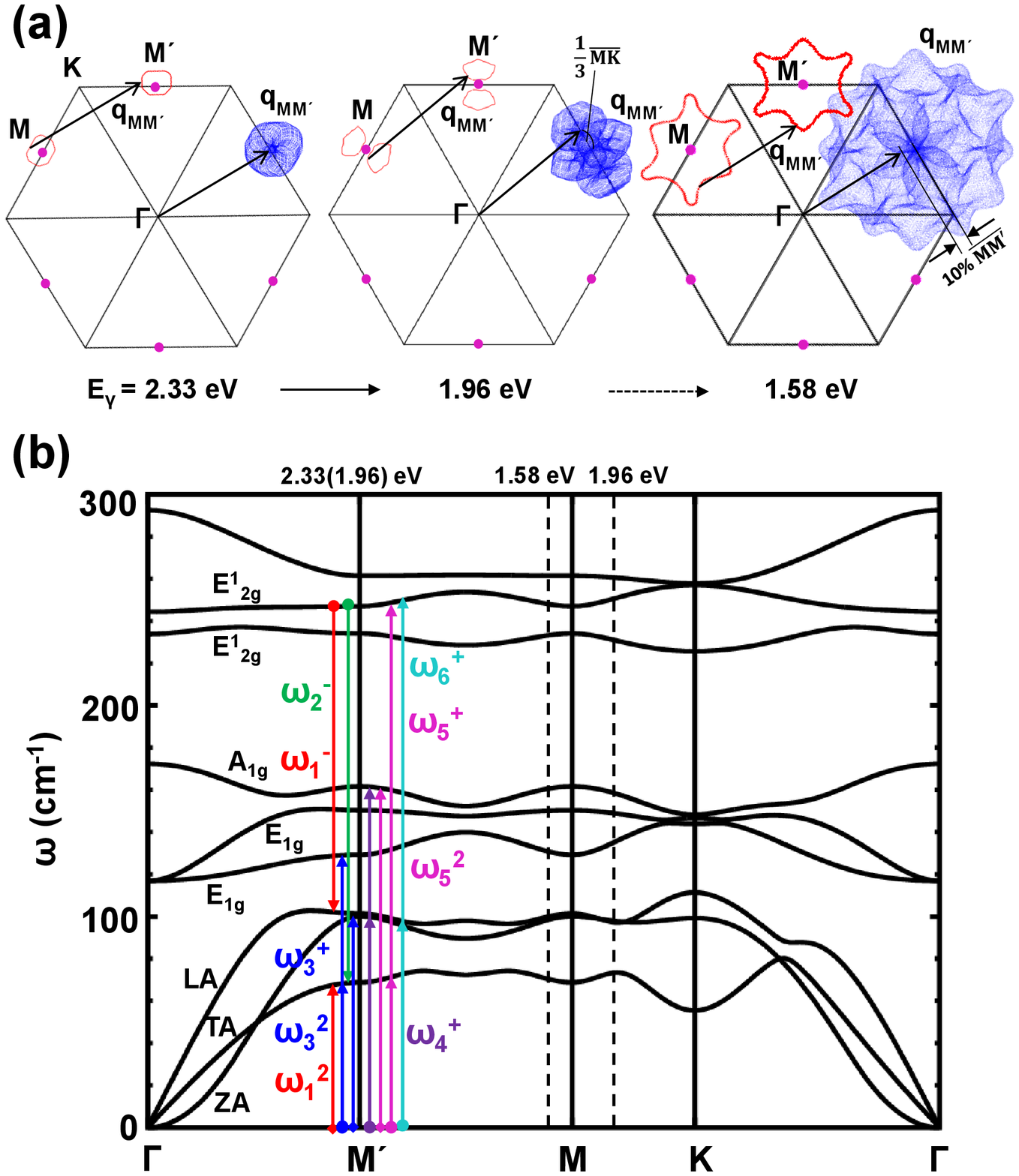}
\caption{(Color online) (a) All possible phonon vectors q$_{\textrm{\textrm{MM}}'}$ coming from the double resonance Raman process due to the laser energies of 2.33, 1.96 and 1.58 eV. The $\vec q_{\textrm{MM}'}$ is equal to either $\overline{\Gamma\textrm{M}}$ (for the laser energy E$_{\gamma}$ = 2.33 and 1.96 eV) or $\sim\frac{1}{3}\overline{\textrm{MK}}$ (for E$_{\gamma}$ = 1.96 eV) or $\sim$10\%$\overline{\textrm{MM}'}$ (for E$_{\gamma}$ = 1.58 eV), as indicated by an arrow away from the $\Gamma$ point. (b) The phonon dispersion relations of a monolayer MoTe$_2$. The second-order phonon modes are analyzed at the M$'$ point for the laser energy E$_{\gamma}$ = 2.33 eV, and then at two k points marked by two vertical dashed lines which are shifted leftward and rightward from the M point for the laser energy E$_{\gamma}$ = 1.58 eV and 1.96 eV, respectively. The red, green, blue, purple, magenta, and cyan vertical arrows are used for modes $\omega_1$, $\omega_2$, $\omega_3$, $\omega_4$, $\omega_5$, and $\omega_6$, respectively. The superscripts "+", "-", and "2" for $\omega$ refer to possible "combination modes" (i.e. $\omega^+_{4}$ = A$_{1g}$(M) + LA(M), $\omega^+_{6}$ = E$^1_{2g}$(M) + LA(M)), "difference modes" (i.e. $\omega^-_{1}$ = E$^1_{2g}$(M) $-$ LA(M), $\omega^-_{2}$ = E$^1_{2g}$(M) $-$ TA(M)) and "overtones" (i.e. $\omega^2_{1}$ = 2TA(M), $\omega^2_{3}$ = 2LA(M)), respectively.}
\label{Fig4}
\end{figure*}

If the laser excitation energy matches the optical transition energy at the M point, we expect that many photo-excited electrons at the M point are scattered to the inequivalent M$'$ or M$''$ points by an inter-valley resonant electron-phonon interaction, as shown in Fig.~\ref{Fig3}(a). The inter-valley electron-phonon process, together with the electron-photon process, gives rise to a double resonance Raman process. Selecting one process from three possible inter-valley scattering of photo-excited carriers (M$\longleftrightarrow$M$'$ or M$\longleftrightarrow$M$''$ or M$'$$\longleftrightarrow$M$''$), we show the schematic view of the double resonance process in Fig.~\ref{Fig3}(b). In Fig.~\ref{Fig3}(b), a photo-excited electron will be scattered from M to M$'$ by emitting a phonon with energy $\hbar \omega_{p1}$ and momentum $-\vec q$, which conserves the energy and momentum of both the electron and the phonon. Then the electron is scattered back to M by emitting the second phonon with energy $\hbar \omega_{p2}$ and momentum $+\vec q$. As analyzed in the Fig.~\ref{Fig3}(a), the momentum $\vec q$ of the phonon can only have a magnitude of $|\overrightarrow{\textrm{MM}'}|$ or $|\overrightarrow{\textrm{MM}''}|$ or $|\overrightarrow{\textrm{M}'\textrm{M}''}|$. When we measure the value of $|$q$|$ from the $\Gamma$ point, the $\vec q$ corresponds to $\overrightarrow{\Gamma \textrm{M}''}$ or $\overrightarrow{\Gamma \textrm{KM}'}$ or $\overrightarrow{\Gamma \textrm{M}}$, respectively. Since $\overrightarrow{\Gamma \textrm{KM}'}$ = $\overrightarrow{\Gamma \textrm{M}'}$ + $\overrightarrow{\Gamma \Gamma}$ $\equiv$ $\overrightarrow{\Gamma \textrm{M}'}$, all phonon modes related to double resonance process correspond to phonon modes at the M point.

\begin{table*}[t]
\caption{Phonon frequencies of the first- and second-order Raman spectra of monolayer MoTe$_2$ are listed for three different laser energies E$_\gamma$. A comparison for first and second-order Raman spectra is given between the experimentally unassigned modes $\omega_i$ (i = 1, 2, ..., 6) and the calculated phonon frequencies both at the M point and near the M point denoted by (off M) whose $\vec q$ lies around $\frac{1}{3}\overline{\textrm{MK}}$ and 10\%$\overline{\textrm{MM}'}$ for E$\gamma$ = 1.96 eV and 1.58 eV, respectively. The consistent dispersion behavior for the second-order Raman spectra between the experiment and the theory are underlined. The frequency is given in units of cm$^{-1}$. NA: not available.}
\begin{tabular}{p{1.0cm}|p{1.60cm}p{1.60cm}p{1.60cm}|p{1.60cm}p{2.10cm}p{1.60cm}|p{4.10cm}}
\hline
\hline
{}&{}&{}&{}&{}&{} &{} &{}\\
{}&\multicolumn{3}{c|}{Experiment}&\multicolumn{3}{c|}{Calculation}&{Phonon assignments}\\
{}&{}&{}&{}&{} &{} &{}&{}\\
\hline
{E$_\gamma$}&{2.33 eV}&{1.96 eV}&{1.58 eV}&{2.33 eV}&{1.96 eV}&{1.58 eV}&{}\\
\hline
{E$_{1g}$} &  {NA }&{NA} &{NA}      &{116.9}&{ }&{ }  &{first-order}\\
{A$_{1g}$}&{171.1}&{171.3}&{170.0}   &{172.2}&{ }&{ }  &{first-order}\\
{E$^1_{2g}$} &{235.8} &{236.2}&{234.1}&{236.9}&{ }&{ } &{first-order}\\
\hline
{}&{}&{}&{}&{(at M)}&{(at M/off M)}&{(off M)}&{}\\
\hline
{$\omega_1$}&{138.2}&{140.7}&{139.0}& {\underline{137.4}}&{{137.4}/\underline{146.6}}&{\underline{140.1}}&{\underline{2TA(M)}} \\
{          }&{     }&{     }&{     }&{\underline{132.8}}&{{132.8}/\underline{133.1}}&{\underline{133.2}}&{or \underline{E$^1_{2g}$(M)-LA(M)}}\\
{$\omega_2$}&{183.7}&{183.6}&{181.5}&{\underline{178.3}} &{\underline{178.3}/{177.2}} &{\underline{177.8}}& {\underline{E$^1_{2g}$(M)-TA(M)}}\\
{$\omega_3$}&{205.5}&{205.3}&{205.9}&{\underline{197.9}} &{\underline{197.9}/{208.2}} &{\underline{200.5}}&{\underline{E$_{1g}$(M)+TA(M)}}\\
{          }&{     }&{     }&{     }&{\underline{202.7}}  &{\underline{202.7}/196.0} &{\underline{200.4}}&{or {\underline{2LA(M)}}}\\
{$\omega_4$}&{$\sim$264.0}&{260-270}&{260-270}&{\underline{263.0}} &{263.0/\underline{256.6}} &{\underline{261.0}}&{\underline{A$_{1g}$(M) + LA(M)}}\\
{$\omega_5$}&{{329.0}}&{{333.2}}&{$\sim$328.0}&{\underline{323.3}} &{323.3/317.2} &{\underline{321.6}}&{2A$_{1g}$(M)}\\
{}&{}&{}&{}&{\underline{315.7}}&{315.7/\underline{323.8}} &{\underline{317.9}}&{or \underline{E$^1_{2g}$(M)+TA(M)}}\\
{$\omega_6$}&{347.4}&{348.4}&{346.5}&{\underline{348.3}}  &{\underline{348.3}/348.5} &{\underline{348.0}}&{\underline{E$^1_{2g}$(M)+LA(M)}}\\
\hline
\hline
\end{tabular}
\label{Table1}
\end{table*}


In order to specify the phonons contributing to the second-order Raman process, we analyzed possible inter-valley phonon vectors $\vec q_{\textrm{MM}'}$ for the three laser energies and calculated the phonon spectra of monolayer MoTe$_2$, as shown in Fig.~\ref{Fig4}. For the laser with an energy of 2.33 eV, we analyze possible $\vec q_{\textrm{MM}'}$ wave vectors between two elliptical equi-energy contour lines (red lines) in the left panel of Fig.~\ref{Fig4}(a). The dark area for $\vec q_{\textrm{MM}'}$ shows that the most probable phonon vectors appear at the M point. Other phonon vectors, such as $\vec q_{\textrm{M}'\textrm{M}''}$ and $\vec q_{\textrm{MM}''}$, are similar to $\vec q_{\textrm{MM}'}$ (not shown). For the 1.96 and 1.58 eV excitation lasers, as already seen in Fig.~\ref{Fig2}(c), the equi-energy contours have a larger area than that for 2.33 eV. Then we expect that $\vec q_{\textrm{MM}'}$ for E$\gamma$ = 1.96 and 1.58 eV arises from the vicinity of the Brillouin zone near the M point, as shown in the central and right panels of Fig.~\ref{Fig4}(a). The relatively dark region indicates that: (1) the $\vec q_{\textrm{MM}'}$ for E$\gamma$ = 1.96 eV appear both at the M point and in the vicinity of the M point (approximately with lengths of $\frac{1}{3}\overline{\textrm{MK}}$) along the MK direction and (2) the $\vec q_{\textrm{MM}'}$ for E$\gamma$ = 1.58 eV appear in the vicinity of the M point along the M$'$M$''$ (or \textrm{MM}$''$) direction approximately with lengths of 10\%$\overline{\textrm{M}'\textrm{M}''}$ (or 6.5\%$\overline{\textrm{MM}''}$) away from the M$''$ point.

In Fig.~\ref{Fig4}(b), we show the calculated phonon dispersion relations of monolayer MoTe$_2$. The symmetry for each optical phonon band\cite{note1}, such as E$_{2g}^1$ and E$_{1g}$, is labeled in Fig.~\ref{Fig4}(b). Here LA, TA and ZA refer to the longitudinal, transverse and out-of-plane acoustic modes, respectively. We assign the second-order Raman modes at the M point for the laser energy of 2.33 eV. The strategy for the assignment is to choose any two phonon modes at the M point as one overtone or (difference) combination mode and to list in Table~\ref{Table1} those that have frequency values within the experimental $\omega_i \pm$ 6 cm$^{-1}$. These assignments are marked in Fig.~\ref{Fig4}(b) by using color arrows. The red, green, blue, purple, magenta, and cyan arrows represent overtone or combination Raman modes of $\omega_i$ (i = 1, 2, ..., 6), respectively. Downward arrows with an a "$-$" superscript represent a two-phonon process with one phonon annihilated and one phonon created to generate "difference combination mode", while upward arrows with the "+" or "2" superscripts are used for combination or overtone modes in which two phonons are created. For example, $\omega_1$ is assigned to either the $\omega^-_1$ (= E$^1_{2g}$(M) - LA(M)) or $\omega^2_1$ (= 2TA(M)), $\omega_2$ to $\omega^-_2$ (= E$^1_{2g}$(M) - TA(M)), $\omega_3$ to either $\omega^2_3$ (= 2LA(M)) or $\omega^+_3$ (= E$_{1g}$(M) + TA(M)), $\omega_4$ to $\omega^+_4$ (= A$_{1g}$(M) +LA(M)), $\omega_5$ to either the $\omega^2_5$ (= 2A$_{1g}$(M)) or $\omega^+_5$ (= E$^1_{2g}$(M) + TA(M)) and $\omega_6$ to the $\omega^+_6$ (= E$^1_{2g}$(M) + LA(M)) mode. Using the same phonon assignment, we obtain the second-order Raman frequencies for E$\gamma$ = 1.96 eV both at the M point and near the M point with a length of $\frac{1}{3}\overline{\textrm{MK}}$ and for 1.58 eV at one point away from the M point with 10\%$\overline{\textrm{M}'\textrm{M}''}$.

\begin{figure}[t]
\includegraphics[width=.70\columnwidth]{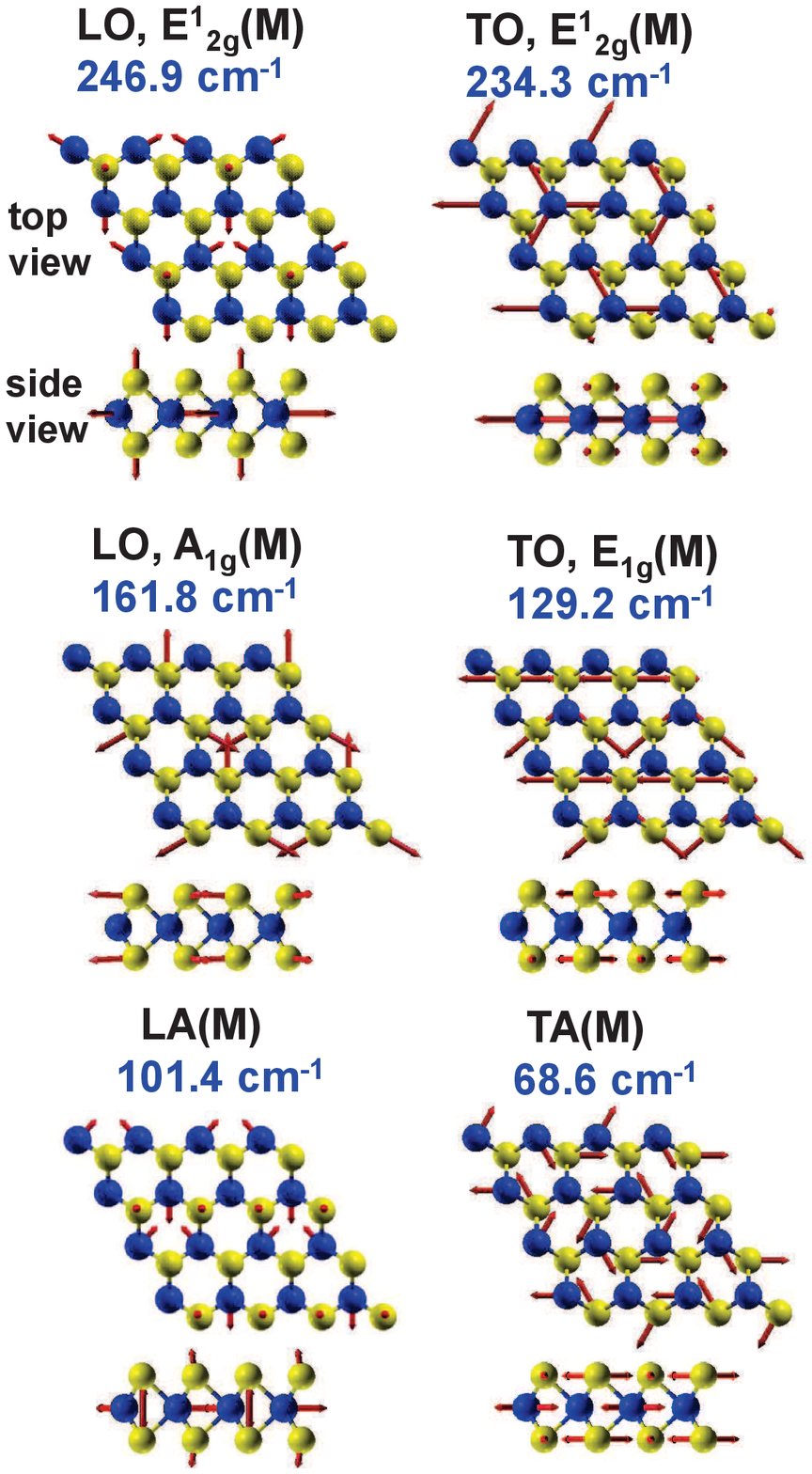}
\caption{(Color online) Visualization of the possible Raman-active modes at the M point for the second-order process suggested in Figure 4. Both the top and side views are given and red arrows represent the vibrational directions and amplitudes of the motion of individual atoms.}
\label{Fig5}
\end{figure}

In Table~\ref{Table1}, we list phonon frequencies of the first-order (upper) and the second-order (lower) Raman spectra for both the experiment and the calculation and also the phonon assignments. The first-order experimental Raman spectra (A$_{1g}$ and E$^1_{2g}$) do not change the frequency for the laser energies (2.33 and 1.96 eV) within the experimental error (1 cm$^{-1}$) and agree with the calculations. The small frequency downshift of the first order spectra (1.7 cm$^{-1}$ at most) from 2.33 eV to 1.58 eV laser excitation is checked to be not from the laser-induced thermal effect as discussed in the Supplemental Material. Although we did not specify the origin of the downshift precisely, the origin of the change for the first order can be neglected for simplicity. The experimental second-order peak positions of $\omega_i$ are either upshifted or downshifted with the laser energy E$_\gamma$ by a small magnitude of about 1-3 cm$^{-1}$. It is pointed out that the frequency of $\omega_1$ is upshifted from 2.33 eV to 1.58 eV even though the both the first-order spectra are downshifted. This can be understood by double resonance Raman theory in which the resonance phonon frequency changes along the phonon dispersion near the M point. In the right of Table~\ref{Table1}, the second-order phonon frequencies for 2.33 and 1.58 eV laser excitations are obtained at $\vec q_{\textrm{MM}'}$ = $\overrightarrow{\Gamma\textrm{M}}$ (at M) and $\sim$10\%$\overline{\textrm{MM}'}$ (denoted by "off M"), respectively. For E$_\gamma$ = 1.96 eV, the phonon assignments are obtained either at $\vec q_{\textrm{MM}'}$ = $\overrightarrow{\Gamma\textrm{M}}$ (at M) or $\sim\frac{1}{3}\overline{\textrm{MK}}$ (off M).

These assignments can be further confirmed by the observed relative Raman intensity and the phonon dispersion relation. First, in Fig.~\ref{Fig1}(b), a larger relative intensity of Raman spectra $\omega_i$ can be found at E$\gamma$ = 2.33 eV than at 1.58 eV. This supports the assignments at the M point and at $\vec q_{\textrm{MM}'}$ $\sim$10\%$\overline{\textrm{MM}'}$ for E$\gamma$ = 2.33 eV and 1.58 eV, respectively, since phonon density of states (PhDOS) at the M point is larger than that at $\vec q_{\textrm{MM}'}$ $\sim$10\%$\overline{\textrm{MM}'}$ as shown in Fig.~\ref{Fig4}(b). Similarly, for E$\gamma$ = 1.96 eV,
a relatively large (small) intensity of the Raman spectra $\omega_2$, $\omega_3$ and $\omega_6$ ($\omega_1$, $\omega_4$ and $\omega_5$) in Fig.~\ref{Fig1}(b) may arise from the large (small) PhDOS at the M point (off the M point). It is therefore reasonable to assign the $\omega_2$, $\omega_3$ and $\omega_6$ ($\omega_1$, $\omega_4$ and $\omega_5$) at the M point (off the M point), as underlined in Table~\ref{Table1}. Second, it is noted that the experimental spectra of $\omega_4$ (at the three laser energies) are too weak to determine the frequency values. The $\omega_4$ is assigned as a combination mode related to the A$_{1g}$(M) as listed in Table~\ref{Table1}. Then it is interesting to discuss why the $\omega_4$ becomes weaker and broader when the first-order A$_{1g}$($\Gamma$) mode becomes stronger by changing the laser excitation energy from 2.33 eV to 1.96 eV/1.58 eV. We expect that the photo-excited electron may choose either an intra-valley scattering process (A$_{1g}$($\Gamma$)) or an inter-valley scattering process (A$_{1g}$(M)), so that A$_{1g}$(M) competes with A$_{1g}$($\Gamma$), resulting in the alternative Raman intensity of A$_{1g}$($\Gamma$) and $\omega_4$ in the experiment. The $\omega_4$ (A$_{1g}$(M)) has no VHS of PhDOS compared with all the other $\omega$'s at the M point as seen from Fig.~\ref{Fig4}(b) which is relevant to the weak intensity of $\omega_4$, too.

In Fig.~\ref{Fig5}, we show the normal modes of the phonons at the M point in which the phonon vibration amplitude is visualized in the 2$\times$2 supercell. As seen from Fig.~\ref{Fig4}(b), the two groups of optical phonon modes (E$_{2g}^1$ and (E$_{1g}$ or A$_{1g}$)) are well separated by about 50 cm$^{-1}$. The visualization of these M point phonon modes shows that the LO and TO E$_{2g}$(M) modes (230$\sim$300 cm$^{-1}$) and the LA(M) and TA(M) modes (60$\sim$100 cm$^{-1}$) are bond-stretching modes of the Mo-Te bond. The LO A$_{1g}$(M) and  TO E$_{1g}$(M) modes (120$\sim$180 cm$^{-1}$) arises from the relative motion of two Te atoms occupying the same sublattice. The six vibrational modes in Fig.~\ref{Fig5} are Raman active, since the volume or the area of the supercell under vibration gives non-zero deformation potential. Thus all the modes can contribute to the 2nd-order overtone, combination and difference modes. Furthermore, the vibrational magnitude of the TA modes is no less than that of the LA modes, suggesting that the TA modes may induce as large deformation potential as the LA modes in the electron-phonon interaction. Usually, the TA modes around the $\Gamma$ point do not contribute to the electron-phonon interaction. Nevertheless, this seems not to be the case at the M point.

So far we have only considered the electron-involved double resonance process for MoTe$_2$ and did not discuss the hole-involved double resonance process. As Venezuela et al. \cite{Venezuela11} pointed out, the enhancement by the double resonance is strong in graphene for one-electron and one-hole scattering, since the energy denominators of the Raman intensity formula becomes zero for the symmetric $\pi$ and $\pi^*$ bands of graphene (triple resonance). In the case of TMDs, however, a similar effect is not expected since the conduction band and the valence band are not symmetric around the Fermi energy \cite{Wang14}. Since we don't have a program to calculate electron-phonon matrix element by first-principles calculations, we can't calculate Raman intensity for each double resonance process,\cite{Venezuela11} which will be a future work.
\begin{table}[t]
\caption{Laser energies that are used in the Raman spectra of TMDs. E$_g$(K), E$_g$(M) are, respectively, the optical transition energies at the K and M point, which are obtained by LDA calculations. Detail is discussed in the text.}
\begin{tabular}{lllll}
\hline
{TMDs} & {laser(eV)} &{E$_g$(K)}  &  {E$_g$(M)} \\
\hline
{MoS$_2$}\cite{Chen74,Stacy85} & {1.83-2.41} & {1.7}  &  {$>$3.0} \\

{WS$_2$}\cite{Sourisseau89,Sourisseau1991,Berkdemir13} &{1.83-2.41}  & {1.9}  &  {$>$3.0} \\

{WSe$_2$}\cite{Zhao13,Terrones14} & {2.41}  & {1.7}  &  {$>$3.0} \\

{MoTe$_2$}$^{\emph{this work}}$ & {1.96, 2.33}  & {1.1}  &  {$\sim$1.8}  \\
\hline
\end{tabular}
\label{Table2}
\end{table}

Finally, we briefly comment on the exciton effect on the double resonance Raman spectra of MoTe$_2$. Since the excitonic states appear at the direct energy gap at the K point of the BZ\cite{Chernikov14,Ye14}, we expect that the excitonic energy is around 1.1 eV for MoTe$_2$, which is much smaller than the three laser energies we used (1.58 - 2.33 eV). As far as we consider only MoTe$_2$, we can safely neglect the exciton effect at the K point. However, in the case of other TMDs (MoS$_2$\cite{Stacy85}, WS$_2$\cite{Sourisseau89,Sourisseau1991,Berkdemir13,ws2ki}, WSe$_2$\cite{Zhao13,Terrones14}) energy gaps at the K point (E$_g$(K)) lie in the range
of 1.7-1.9 eV as shown in Table~\ref{Table2}. The laser energies used in the experiments lie in a range from 1.83 to 2.41 eV. Thus the exciton effect needs to be considered for some laser energies. Nevertheless, when we discuss the M-point phonon of the other TMDs, we expect that the second-order Raman spectra do not appear, since the optical transition energies at the M point is larger than 3.0 eV. It would be interesting if ultra-violet laser could be used for detecting the M-point phonon modes for other TMDs, which will be a future work. In the case of MoTe$_2$, on the other hand, since E$_g$(M) $\sim$ 1.8 eV, we expect the double resonance Raman process for conventional laser energies. Although we did not consider the exciton effect at the M point explicitly as K.F. Mak \textit{et al.}\cite{Mak11} did in graphene, we believe that the present discussion of the double resonance Raman effect should work well, because only optical transition matrix elements that are common in the resonant Raman process are enhanced by the exciton effect, and the relative Raman intensity thus obtained do not change much for the M point. \cite{Jiang07a,Jiang07b}

\section{Conclusion}

In summary, we have investigated the second-order Raman spectra of MoTe$_2$ systems, based on an {\em ab initio} density functional calculation, and compared with the experimental Raman spectroscopy. The calculated electronic bands, the Van Hove singularities in the density of states, and the absorption spectra indicate a strong optical absorption at the M point in the BZ of MoTe$_2$. Using double resonance Raman theory, combination/difference or overtone mode of phonons at the M point are assigned to identify the experimental Raman modes of $\omega_i$ (i = 1, 2, ..., 6). This study facilitates a better understanding of the electron-phonon interaction and the second-order Raman process in TMD systems.

\section*{Acknowledgments}
M.Y. and K.T. sincerely acknowledge Prof. Meiyan Ni and Dr. Katsunori Wakabayashi for critical suggestions on experiments. T.Y. and Z.D.Z acknowledge the NSFC under Grant Nos. 11004201, 51331006 and the IMR SYNL-Young Merit Scholars for financial support. R.S. acknowledges MEXT Grants Nos. 25286005 and 25107005. M.S.D. acknowledges NSF-DMR Grant No. 10-04147. K.U. and K.T. acknowledge MEXT Grant No. 25107004. Calculations were performed in the HPC cluster at the Institute of Metal Research in Shenyang, China.


\begin{thebibliography}{61}
\bibitem[{not()}]{note1}
\bibinfo{note}{The phonon modes E$^{''}$, E$^{'}$ and A$_1^{'}$ of monolayer
  and related Raman-active modes of few-layer MoTe$_2$ are labeled as E$_{1g}$,
  E$^{1}_{2g}$ and A$_{1g}$ of bulk MoTe$_2$ for discussion convenience.}
\end{thebibliography}

\end{document}